\documentclass[manuscript]{aastex63}

\usepackage{amsmath,amssymb,bm}
 \usepackage{color}
 \definecolor{ForrestGreen}{rgb}{0.133,0.545,0.133}
 \definecolor{DarkGreen}{rgb}{0.0,0.45,0.0}
  \newcommand{\B}[1]{{\color{blue}{\textbf{#1}}}} \newcommand{\R}[1]{{\color{red}{\textbf{#1}}}} \newcommand{\G}[1]{{\color{DarkGreen}{\textbf{#1}}}} \newcommand{\M}[1]{{\color{magenta}{\textbf{#1}}}}
\received{Oct 28, 2020}
\revised{Sep 6, 2021}
\accepted{xxx x, xxxx}

\submitjournal{ApJ}

\shortauthors{Chen et al.}

\begin{document}
\title{Partial Eruption, Confinement, and Twist Buildup and Release of a Double-decker Filament}

\correspondingauthor{Yingna Su}
\email{ynsu@pmo.ac.cn}

\author[0000-0002-2436-0516]{Jialin Chen}
\affiliation{Key Laboratory of Dark Matter and Space Astronomy, Purple Mountain Observatory, CAS, Nanjing, Jiangsu 210023, China}
\affiliation{Department of Astronomy and Space Science, University of Science and Technology of China, Hefei, Anhui 230026, China}

\author[0000-0001-9647-2149]{Yingna Su}
\affiliation{Key Laboratory of Dark Matter and Space Astronomy, Purple Mountain Observatory, CAS, Nanjing, Jiangsu 210023, China}
\affiliation{Department of Astronomy and Space Science, University of Science and Technology of China, Hefei, Anhui 230026, China}

\author[0000-0003-4618-4979]{Rui Liu}
\affiliation{CAS Key Laboratory of Geospace Environment, Department of Geophysics and Planetary Sciences, University of Science and Technology of China, Hefei, Anhui 230026, China}
\affiliation{Collaborative Innovation Center for Astronautical Science and Technology, Hefei, Anhui 230026, China}

\author[0000-0002-5740-8803]{Bernhard, Kliem}
\affiliation{Institute of Physics and Astronomy, University of Potsdam, D-14476 Potsdam, Germany}

\author[0000-0003-4078-2265]{Qingmin Zhang}
\affiliation{Key Laboratory of Dark Matter and Space Astronomy, Purple Mountain Observatory, CAS, Nanjing, Jiangsu 210023, China}

\author[0000-0002-5898-2284]{Haisheng, Ji}
\affiliation{Key Laboratory of Dark Matter and Space Astronomy, Purple Mountain Observatory, CAS, Nanjing, Jiangsu 210023, China}
\affiliation{Department of Astronomy and Space Science, University of Science and Technology of China, Hefei, Anhui 230026, China}

\author[0000-0001-7385-4742]{Tie Liu}
\affiliation{Key Laboratory of Dark Matter and Space Astronomy, Purple Mountain Observatory, CAS, Nanjing, Jiangsu 210023, China}
\affiliation{Department of Astronomy and Space Science, University of Science and Technology of China, Hefei, Anhui 230026, China}


\begin{abstract}

We investigate the failed partial eruption of a filament system in NOAA AR 12104  on 2014 July 5, using multiwavelength EUV, magnetogram, and H$\alpha$ observations,  as well as magnetic field modeling. The filament system consists of two almost co-spatial segments with different end points, both resembling a C shape. Following an ejection and a precursor flare related to flux cancellation, only the upper segment rises and then displays a prominent twisted structure, while rolling over toward its footpoints. The lower segment remains undisturbed, indicating that the system possesses a double-decker structure. The erupted segment ends up with a reverse-C shape, with material draining toward its footpoints, while losing its twist. Using the flux rope insertion method, we construct a model of the source region that qualitatively reproduces key elements of the observed evolution. At the eruption onset, the model consists of a flux rope atop a flux bundle with negligible twist, which is consistent with the observational interpretation that the filament possesses a double-decker structure. The flux rope reaches the critical height of the torus instability during its initial relaxation, while the lower flux bundle remains in stable equilibrium. The eruption terminates when the flux rope reaches a dome-shaped quasi-separatrix layer that is reminiscent of a magnetic fan surface, although no magnetic null is found. The flux rope is destroyed by reconnection with the confining overlying flux above the dome, transferring its twist in the process.

\end{abstract} 

\section{Introduction} 

Filaments, also called prominences when they appear above the solar limb, are a cool plasma material that is suspended in the solar corona, supported and confined by the coronal magnetic field. Filaments are closely related to the most violent solar activities, i.e., filament eruptions, solar flares, and coronal mass ejections (CMEs). In many cases, but not all, a filament is identified to exist in the core of a CME \citep{hundhausen1999}. Thanks to high-resolution observations, a growing number of recent studies suggest that filaments sometimes consist of a double-decker or even multistranded structure before erupting \citep{liu2012b, lishangwei2017, awasthi2018, su2018, Pan2021}. If two filaments are associated with the same magnetic polarity inversion line (PIL) and separated in height, such a system is called a double-decker structure \citep{liu2012b}. The two filaments may erupt together or sequentially \citep[e.g.,][]{zhu2015}, or in some cases, the upper one erupts outward while the lower one stays almost unchanged \citep{liu2012b, zhu2014, zheng2019}. 

Solar filament eruptions have been classified into three categories, i.e., full, partial, and failed eruptions. In a full eruption, the entire magnetic structure erupts with most of the preeruptive filament mass \citep{Tang1986, Plunkett2000}. While in a failed filament eruption, the filament initially shows an eruptive-like rise, and then falls back toward the Sun after reaching a maximum height \citep[e.g., ][]{ji2003, alexander2006, Liul2018}. Only part of the filament structure has erupted in a partial filament eruption \citep{Tang1986, Gilbert2000, Gibson2002, Pevtsov2002}. For a detailed classification of filament eruptions, we refer to \citet{gilbert2007}. The full and partial eruptions are generally larger in scale and more energetic, while the failed eruptions are often restricted to a local area; though, the association with GOES M or X-class flares has also been reported in some cases.
  
Solar eruptions are suggested to be initiated by ideal  MHD instabilities or, alternatively, by nonideal fast magnetic reconnection. The torus instability and the helical kink instability are two ideal MHD instabilities of a flux rope or current channel often reported as trigger mechanisms for eruptions. In a general definition \citep[][Ch. 2.9.1]{Priest2014}, a magnetic flux rope is a twisted flux tube, with the boundary of the flux tube being given by a set of field lines that intersect a simple closed curve. Torus instability occurs when the overlying field of a filament attenuates fast enough with height \citep{kliem2006}. The filament will then rise faster and faster, and it may eventually erupt into the interplanetary space. The threshold for the onset of torus instability is defined in terms of the decay index of the external poloidal field at the position of the current channel, i.e., $n=-\partial \ln {B_\mathrm{ex}}$/${\partial \ln {h}}$. The canonical values of the critical decay index are 1.5 and 1.0 for toroidal and straight current channels, respectively \citep{Bateman1978, vanTend1978}. In theoretical calculations and numerical simulations by \citet{kliem2006}, \citet{Fan2007}, \citet{aulanier2010a}, and \citet{demoulin2010}, the critical decay index has been found to lie between 1.0 and 2.0. The helical kink instability will occur if the magnetic field supporting the filament accumulates a sufficiently strong twist, e.g., $\Phi_\mathrm{cr}=2.5\pi$ for a uniformly twisted, line-tied flux rope \citep{hood1979} and $\Phi_\mathrm{cr}\sim3.5\pi$ for more realistic flux rope models as expected on the Sun \citep{fan2003, fan2004, torok2004}. Exceeding this threshold, the flux rope will erupt, and part of its twist will transfer into writhe. The instability then saturates quickly; however, when it lifts the rope above the critical height for the onset of the torus instability, a full eruption can result. 

In a \textit{confined} or \textit{failed} eruption, the filament first rises, then halts, and finally falls back toward the Sun \citep[e.g.,][]{ji2003}. Several mechanisms can explain the confinement of the eruption; see \citet{hassanin2016} for a detailed review. The two most  frequently inferred  mechanisms are the following. In eruption models based on ideal MHD instability \citep{vanTend1978}, the eruption fails when the condition for the torus instability is not met in a height range at some distance above the initial destabilization of the flux \citep[e.g.,][]{torok2005, Guo2010}. On the other hand, a failed filament eruption can also occur due to the magnetic reconnection between the erupting and overlying flux, which is facilitated by the existence of a coronal magnetic null point spanning a dome-shaped magnetic fan surface above the eruption site \citep[e.g.,][]{wang2012}. 
 
In this work, we study the failed eruption of a filament system with double-decker structure. The activities which occurred near the northwestern footpoint of the lower filament are found to be crucial in the evolution of the system imbalance and the eruption of the upper filament. We analyze the morphology and evolution of this filament system in detail, in order to figure out what initiates the eruption and why the eruption fails. This paper is organized as follows. Section \ref{obs} presents the data set and instruments. Observational analysis and results are presented in Section \ref{data}. Magnetic modeling of the double-decker system and evolution of the eruption are presented in Section \ref{modeling}. We discuss the initiation and confining mechanisms of the failed eruption in Section \ref{discussion}. A summary and the conclusions are given in Section \ref{summary}. 

\section{Data Set and Instruments} \label{obs}

A failed eruption of a double-decker filament system occurs in NOAA AR 12104 on 2014 July 5. This eruption is associated with a C2.5-class solar flare, but does not produce a CME. The Solar Dynamics Observatory (SDO) satellite \citep{pesnell2012} observes the whole eruption. EUV and magnetic field observations are provided, respectively, by the Atmospheric Imaging Assembly \citep[AIA;][]{lemen2012} and the Helioseismic and Magnetic Imager \citep[HMI;][]{schou2012} onboard SDO. The AIA images have a temporal resolution of 12 s, and a pixel size of 0$\arcsec$.6. The HMI line of sight (LOS) and vector magnetograms have a temporal resolution of 45 s and 720 s, respectively, and their pixel size is 0$\arcsec$.5. The H$\alpha$ images are provided by the Global Oscillation Network Group (GONG). The spatial and temporal resolutions are about 1$\arcsec$ per pixel and 1 minute, respectively.

\section{Observations} \label{data}

\subsection{Activities Prior to the Eruption}

The GOES light curve of the event is presented in Figure \ref{Figure1}(a). Between about 4 and 1.5 hr before the eruption, filament oscillations occur then decay (Figure \ref{Figure1}(b)). This is followed by the C1.3-class precursor flare. Both of the filament oscillations and the precursor flare are closely related to the intermittent, bright ejections from the northwestern footpoint of the lower filament, and the ejections are associated with flux convergence and cancellation in the photosphere. These series of preflare activities are presented in Figures \ref{Figure1}--\ref{Figure2} and described in detail in the following. 

\subsubsection{Intermittent Bright Ejections}\label{sss:intermittent}

Prior to the eruption, both filament segments in the double-decker structure exhibit a C shape, aligned along the same PIL in the outer (northeastern) part of the active region, and largely overlap each other (Figures \ref{Figure1}(c)--(d)). However, the northwestern footpoints of the two filaments show a clear separation. The location and shape of the upper filament change dramatically during the eruption, as shown in Figures \ref{Figure1}(e)--(f). It is not possible to determine how much the filaments separate in height before and after the eruption, because data from a side view are not available. However, there is evidence suggesting the independence of the two filaments already before the eruption, in addition to the separation of their northwestern footpoints. Bright material is repeatedly ejected along the lower filament from its northwestern footpoint toward the northeast, selected examples of which are marked with white arrows in Figures~\ref{Figure1}(g)--(j). While the lower filament is disturbed by such intermittent ejections,  the upper filament remains stable until the onset of the eruption, suggesting that the upper filament is sufficiently separated from the lower one, i.e. a double-decker configuration. Further evidence is provided by signatures of reconnection in the space between the filaments at the onset of the eruption (see Section~\ref{ss:main_eruption}).

\subsubsection{Oscillations and Slow Displacement of the Upper Filament}

The upper filament begins to oscillate longitudinally after the bright ejection at 18:30 UT, which might have triggered the oscillation. The time--distance plot in Figure \ref{Figure1}(b) shows that the oscillation lasts at least five periods, and the oscillation periods are not uniform, ranging from 30 to 40 minutes. The nonuniformity of the oscillations may be due to disturbances by the intermittent ejections. As soon as the oscillations totally disappear at about 21:10~UT (about an hour before the C1.3 precursor flare), a slow westward displacement of the upper filament becomes apparent, suggesting the onset of a slow-rise phase. The projected velocity is estimated from Figure~\ref{Figure1}(b) to be $\approx$37~km\,s$^{-1}$.

\subsubsection{Precursor Flare}\label{ss:precursor} 

\begin{figure*}
    \centering
    \resizebox{0.9\hsize}{!}{\includegraphics{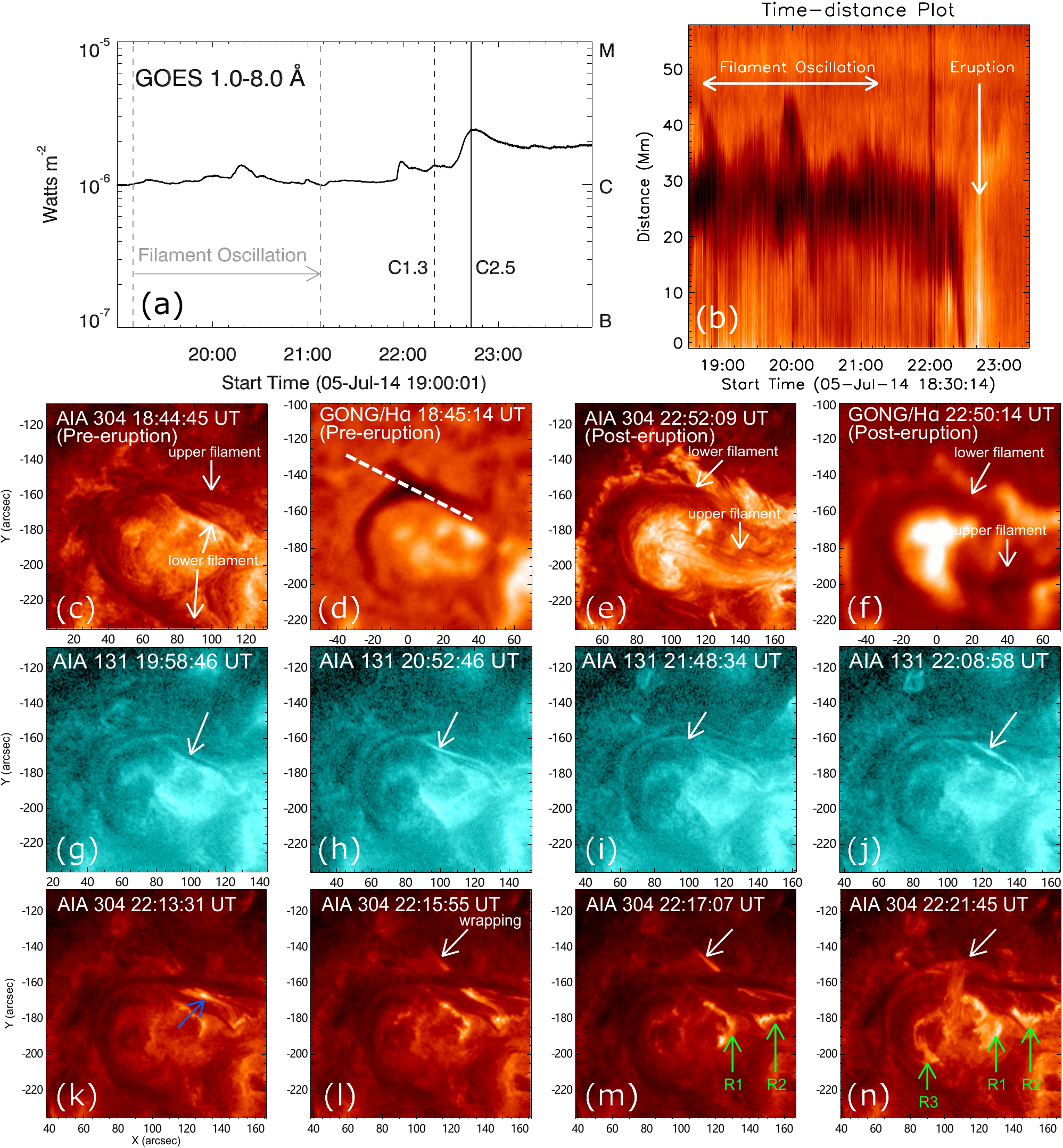}}
    \caption{Overview of the activities prior to the main flare. (a) GOES $1-8${\AA} soft X-ray light curve from 19:00 UT to 23:59 UT. The gray dashed lines refer to the start (19:10 UT) and end (21:08 UT) time of the filament oscillation. The black dashed and solid lines represent the peak time of the precursor and main flares. (b) Time--distance plot of the filament oscillation. (c)--(f) Images of the double-decker filaments (white arrows) before and after the eruption in $304$ {\AA} (AIA) and H${\alpha}$ (GONG). The white dashed line in panel (d) shows the position of the selected slit for the plot in panel (b). (g)--(j) Intermittent bright ejections (white arrows) from the northwest footpoint of the lower filament towards the northeast, observed in 131 {\AA} (AIA) before the eruption. (k)--(n) Evolution of the precursor flare observed in AIA $304$ {\AA}.}
    \label{Figure1}
\end{figure*}

After the filament oscillation, intermittent bright ejections are still ongoing along the lower filament. A particularly strong ejection around 22:05 UT from the northwestern footpoint of the lower filament coincides with the onset of the C1.3 precursor flare, which peaks at 22:20 UT, as shown in Figures \ref{Figure1}(k)--(n). The flare begins with the appearance of two bright flare ribbons (R1 and R2), located on the two sides of the lower filament, at 22:12 UT. The lower filament does not change significantly, whereas the upper filament continues its slow westward displacement. The bright ejection clearly moves under the upper filament to the outer side of the filament channel and largely parallel to the lower filament in this northern leg of the C-shaped structure. After reaching the outer side of the filament channel, the bright ejected material rises and wraps around the upper filament (white arrows in Figures \ref{Figure1}(i)--(n)). This provides strong evidence that the upper filament is embedded in twisted flux. This could be a flux rope or a flux rope in formation, but not a magnetic arcade. The observed upward turn of the ejected material can only be enforced by the magnetic field. Due to the very small values of the plasma beta with a low height in active regions \citep{Gary2001}, moving plasma must usually follow the existing field direction, which here reveals left-handed twist. From its position above the upper filament, the ejected material immediately begins to drain toward the central region surrounded by the filaments, which suggests that the flux rope embedding the upper filament is not yet fully coherent. Associated with the curved ejection is the brightening of another ribbon (R3 in Figure~\ref{Figure1}(n)). The ribbon R3 forms at the periphery of a simultaneously developing dimming area, which it fully encloses from 22:17~UT (Figure~\ref{Figure1}(n)). The dimming suggests that an expanding flux rope is rooted in this area. This is also supported by the draining of the ejected material toward this ribbon: the ejected material appears to be guided by the outer layers of the magnetic flux holding the upper filament.

\subsubsection{Evolution of the Photospheric Magnetic Field}\label{sss:cancellation}

\begin{figure*}
\begin{interactive}{animation}{Figure2-online-animation.mp4}
    \centering
    \resizebox{\hsize}{!}{\includegraphics{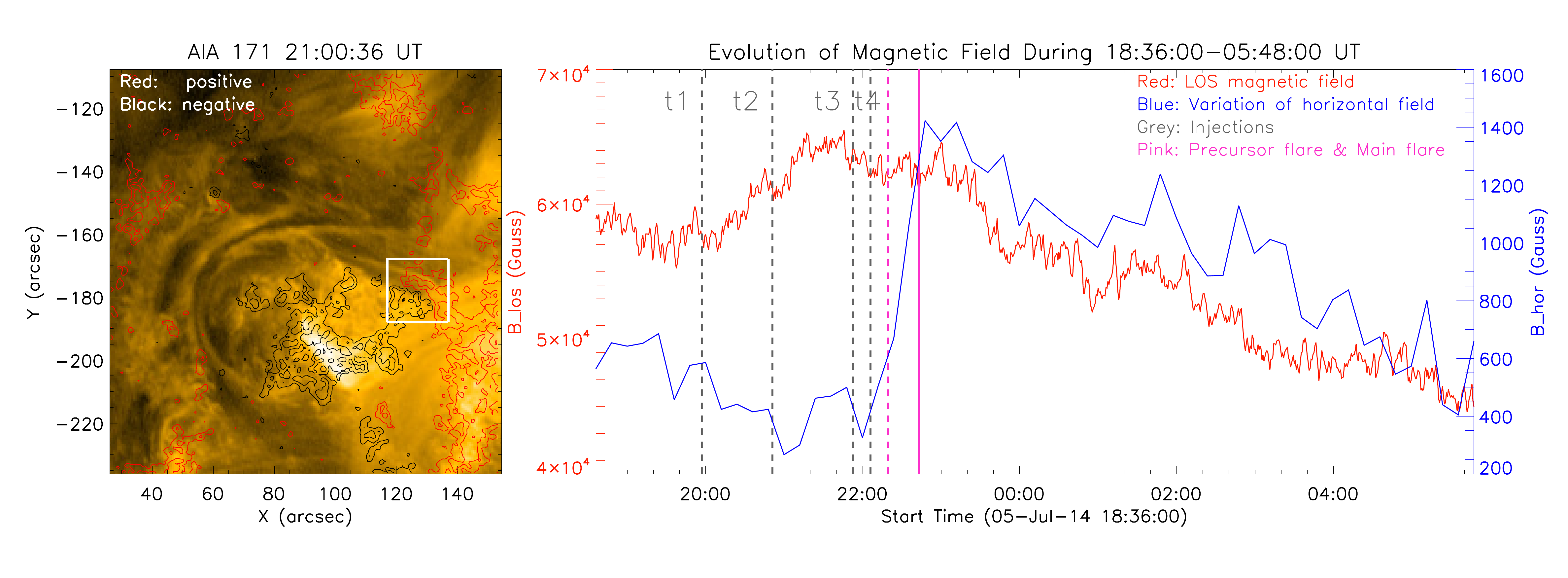}}
\end{interactive}
    \caption{Left panel: AIA 171~{\AA} image overlaid with red (positive) and black (negative) contours representing the LOS photospheric magnetic fields observed by HMI. Right panel: temporal evolution of the total unsigned LOS magnetic field strength (red) and the average horizontal magnetic field strength (blue) within the region enclosed by the white box in the left panel. The pink lines mark the peak times of the precursor flare (22:20 UT) and the main flare (22:43 UT), respectively. The gray dashed lines refer to the time of the intermittent ejections shown in the third row of Figure \ref{Figure1}. An animation of the AIA 171~{\AA} images overlaid with  red (positive) and black (negative) contours representing the LOS photospheric magnetic fields observed by HMI with the same field of view as the left panel, covering the time interval from 18:29 UT on July 5 to 05:58 UT on July 6, is available. The real time duration of the animation is 15 s.\\(An animation of this figure is available.)}
    \label{Figure2}
\end{figure*}

To understand the cause of these activities, we examine the evolution of the photospheric magnetic field observed by SDO/HMI as shown in Figure \ref{Figure2} and the associated animation. The left panel of Figure \ref{Figure2} shows an 171 {\AA} image taken by AIA overlaid with red and black contours representing positive and negative polarity in the magnetogram, respectively. The right panel of Figure \ref{Figure2} presents the temporal evolution of the total unsigned magnetic field strength (red) and the averaged horizontal magnetic field (blue) within the region enclosed by the white box in the left panel of Figure \ref{Figure2}. The pink dashed and solid lines mark the peak times of the precursor flare (22:20 UT) and main flare (22:43 UT), respectively. The gray dashed lines refer to the time of each bright ejection identified in the EUV channels and shown in Figures \ref{Figure1}(g)--(j). Prior to the eruption, two small magnetic polarities are observed to gradually converge and cancel each other at the source region of the bright ejections as enclosed by the white box. The total unsigned magnetic flux within this source region first increases then begins to decrease about 10 minutes before the precursor flare. These changes are due to flux convergence into the selected area and subsequent flux cancellation. The horizontal component of the magnetic field within this source region increases rapidly after the peak of the precursor flare, and then gradually decreases as the main flare and the total unsigned magnetic flux decay. An increase of the horizontal magnetic field component during or after eruptions has been reported, e.g., by \citet{wangshuo2012} and \citet{LiuC2018}, and suggested to be a back-reaction of the photosphere and inner part of the Sun to the flares or CMEs. Alternatively, \citet{Barczynski2019} argued that photospheric horizontal fields are enhanced by the reconnection-driven contraction of sheared flare loops in the corona. 

\subsection{Failed Filament Eruption and Main Flare}\label{ss:main_eruption}

\begin{figure*}
\begin{interactive}{animation}{Figure3-online-animation.mp4}
    \centering
    \resizebox{\hsize}{!}{\includegraphics{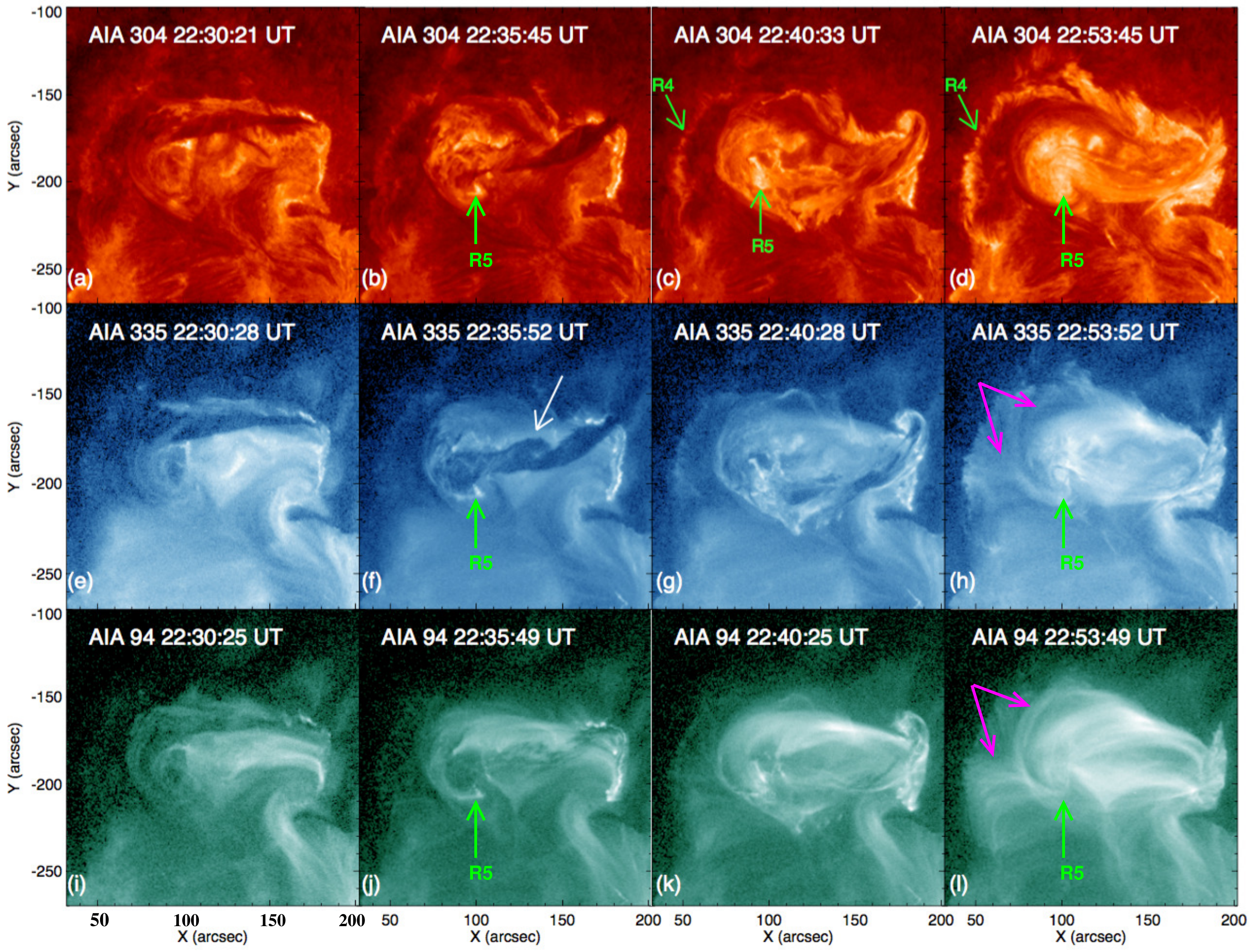}}
\end{interactive}   
    \caption{Twisted internal structure of the erupting filament and release of twist during the main flare. Four snapshots in $304$, $335$, and $94$ {\AA} taken by AIA are presented in the top, middle, and bottom rows, respectively. First column: the eruption begins and the upper filament starts to rise. Second column: the erupting upper filament nearly reaches its maximum height. A clearly twisted structure (white arrow) is best seen in AIA $335$ {\AA}. Third column: the erupted filament untwists during its draining from the terminal position at the west side of the C-shaped filament channel. Fourth column: after the eruption, the upper filament evolves into a much less twisted structure. The C-shaped (outer) and inner ribbons of the main flare are marked as R4 and R5, respectively, and the corresponding flare loops are marked with pink arrows in the bottom row.  An animation of the AIA $304$, $335$, and $94$ {\AA} images with the same field of view as the first column of this figure, covering the time interval from 22:05 UT to 23:00 UT, is available. The real time duration of the animation is 4 s.\\(An animation of this figure is available.)}
    \label{Figure3}
\end{figure*}

Immediately after the onset of the  precursor flare, the bright ejected material wraps around the upper filament, and then drains toward the central region surrounded by the filaments (see bottom row of Figure \ref{Figure1}). Simultaneously (at $\sim$22:15~UT) the rise of the upper filament accelerates, associated with a C2.5 flare that starts during the decay phase of the precursor and peaks at 22:43 UT; see Figure~\ref{Figure3} and the associated animation. Again, the lower filament remains nearly unchanged. By $\sim$22:20~UT, several brightened threads appear along the northern leg of the C-shaped structure, and then wrap about the erupting upper filament in a left-handed sense similar to the bright ejection, and also pass under the filament, which excludes the possibility that they outline flux overlying the filament channel. Their rapid brightening must result from heating by reconnection, which most likely involves ambient flux in the space between the filaments in the standard manner of a two-ribbon flare. The energy deposition into the two ribbons next to the lower filament also indicates such reconnection, which additionally implies that the two filaments reside in separate magnetic structures already by the onset of the eruption, at least along the northern leg of the C-shaped filament channel. The threads show that the flux added to the upper filament by the reconnection carries twist. When it starts to rise, the filament has a straight and slim C shape with only minimal indications of twist. However, it rolls in a way resembling a twisted ribbon ($\sim$22:30~UT), consisting of at least two major intertwined threads ($\sim$22:35~UT; white arrow in Figure \ref{Figure3}(f)), revealing that the upper filament is also internally twisted. 

From a point early in the rise ($\sim$22:20~UT), the erupting filament also shows a motion in the direction to its footpoints. This rolling motion stops at ~22:50 UT after the filament has acquired a reverse-C shape, and subsequently filament material drains toward the Sun. Many brightenings appear intermittently within and below the erupting filament during this final stage of the eruption (Figures \ref{Figure3}(c) and (g)). The filament threads become more or less straightened,  indicating that the filament loses essentially all of its twist during the draining (Figure \ref{Figure3}(d)). This implies that the erupted flux reconnects with ambient flux, different from the standard 2D flare model. Such reconnection is also suggested by the fact that the draining to the negative-polarity side proceeds toward two separate footpoint areas: mostly toward the same region as during the precursor and also toward the southwestern end region of the filament channel. When the upper filament is close to its maximum height, a flare ribbon at the eastern outer side of the filament channel begins to brighten at 22:37 UT. The ribbon then gradually extends and evolves into a full C-shaped ribbon (i.e., outer ribbon R4) enclosing the filament system; see Figures~\ref{Figure3}(c)--(d). The flare loops (pink arrows in Figures~\ref{Figure3}(h) and (l)) rooted in this ribbon connect to a small cusp-shaped ribbon (R5 in Figures~\ref{Figure3}). Similar to ribbon R3 of the precursor flare, ribbon R5 is part of the strongly variable brightening that develops in the periphery of the dimming area. Thus, both ribbons form in the footpoint area of the erupting upper filament. The flare loops rooted in R5 additionally imply that the erupting flux of the upper filament reconnects. The reconnection involves ambient flux rooted in Ribbon~R4.

\section{Modeling and Results} \label{modeling}

\begin{figure*}
    \centering
    \resizebox{0.6\hsize}{!}{\includegraphics{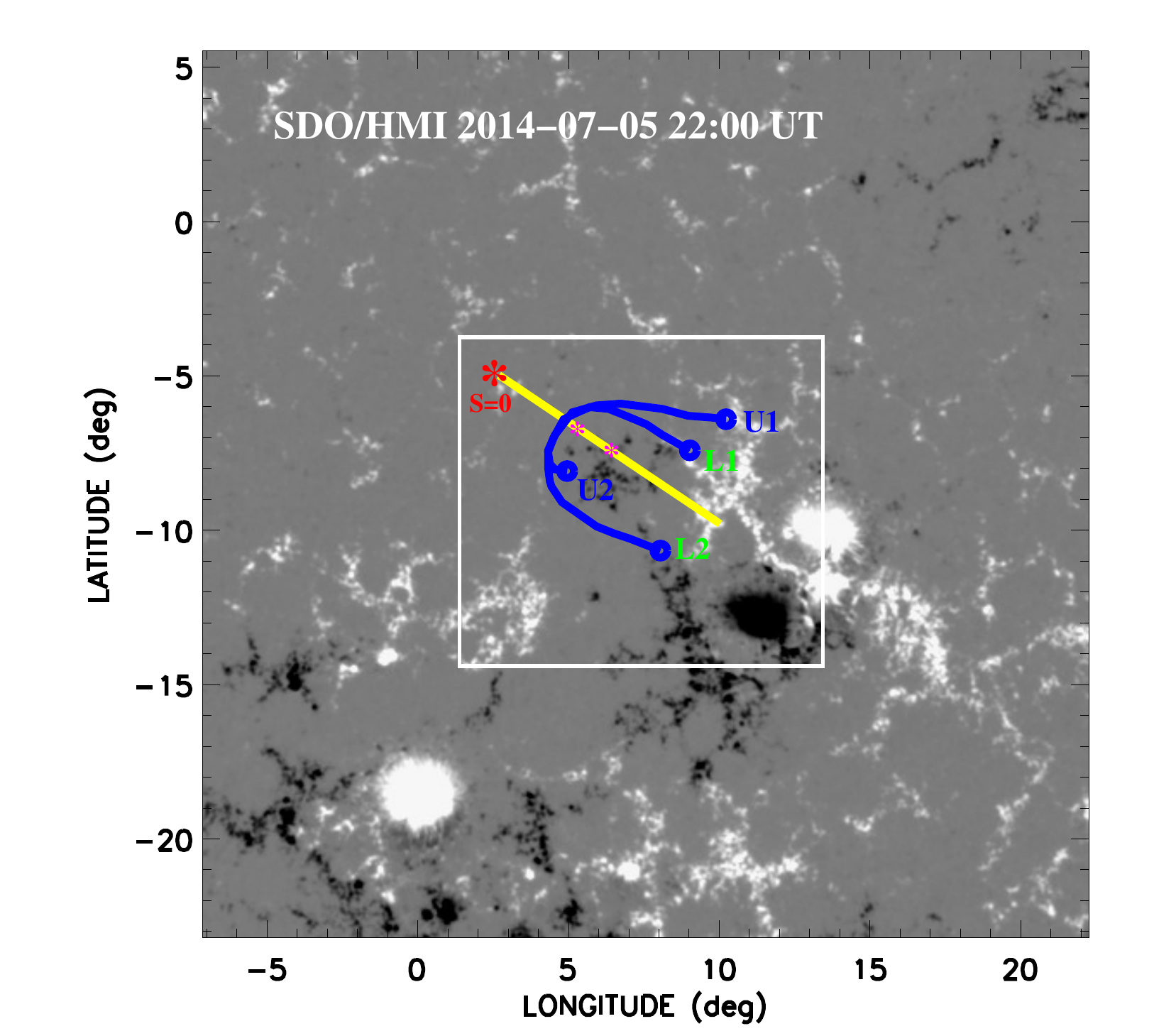}}
    \caption{Longitude--latitude map of the radial component of photospheric magnetic field in the HIRES region of the model taken by SDO/HMI at 22:00 UT. The blue lines ending in circles show the selected filament paths for inserting the flux ropes. The yellow line (start location marked as a red asterisk) refers to the location of the vertical slices in Figures \ref{Figure6}--\ref{Figure8}. Positive and negative polarity are shown in white and black, respectively. The white box marks the approximate field of view (FOV) of Figures~\ref{Figure5}(a)--(b) and the FOV of the left two columns of Figure \ref{Figure6}, and the FOV of Figure \ref{Figure8}(h). The left and right magenta asterisks mark the positions of the solid and dashed curves representing the decay index in Figure \ref{Figure10}.}
    \label{Figure4}
\end{figure*}

To understand the morphology and evolution of the double-decker filament system, we model the magnetic field using the flux rope insertion method developed by \citet{vanBallegooijen2004}. This method is proved to be capable of reproducing nonlinear force-free fields before eruption, as well as unstable magnetic configurations during the early phase of a successful eruption, allowing the stability of filaments to be analyzed \citep[e.g., ][]{su2011, su2015, lishangwei2017, liutie2018}. A detailed description of the methodology and its application can be found in the literature  \cite[and references therein]{vanBallegooijen2004, Su2019}. First, a potential field is computed from the high-resolution (HIRES) HMI magnetogram embedded in a low-resolution HMI global synoptic map. The HIRES region  as shown in Figure \ref{Figure4} spans about 29$^{\circ}$ in both longitude and latitude, and the spatial resolution in the low corona is 0.001 $R_\sun$. The resolution of the global field is about 1$^{\circ}$. Both HIRES and GLOBAL regions extend from the solar surface ($r =$ $R_\sun$) up to a source surface ($r \sim$ 1.7 $R_\sun$) above which the magnetic field is assumed to be radial. Then we create cavities and insert two flux bundles with a double-decker structure along the observed filament paths. The resulting configuration is not in a force-free equilibrium, so the last step is to allow the field to relax through magnetofrictional relaxation \citep{Yang1986, vanBallegooijen2000}. For an unstable model, like the one considered below, relaxation and instability compete in the evolution. Eventually, the instability wins out, which is manifested as a gradual rise of the unstable flux. The magnetofrictional modeling also captures key topological changes but not in the real time scale of the evolution \cite[e.g.,][]{su2011, janvier2016}.

\subsection{Potential Field Model}

\begin{figure*}
    \centering
    \resizebox{\hsize}{!}{\includegraphics{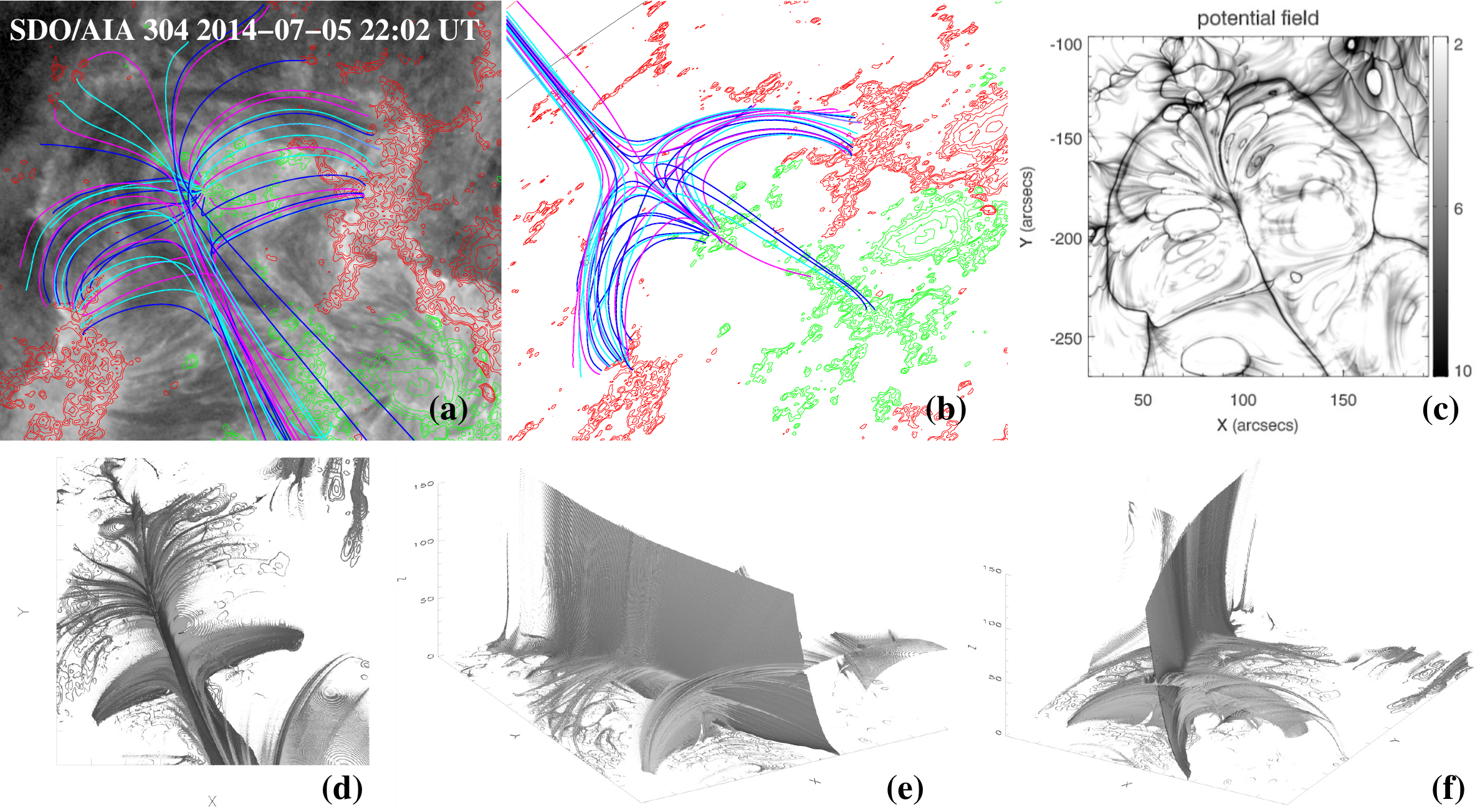}}
    \caption{Results from the potential field model.  (a) Selected field lines from the potential field model overlaid on the AIA image in the $304$ {\AA} channel taken at 22:02 UT. The same field lines viewed from anther angle are shown in (b). The red and green contours represent the positive and negative photospheric magnetic field at 22:00 UT, respectively. (c) Photospheric $\log Q$ map for the potential field model. (d)--(f) Different views of three-dimensional high-$Q$ ($\log Q = 4$) isosurfaces. The field of view of the $X-Y$ plane is similar to that in panel (c), and the Z-axis unit is the cell size of our model (1 cell $\sim{0.7}$ Mm).} 
    \label{Figure5}
\end{figure*}

We investigate the structure of the potential field by computing the squashing factor $Q$, which is a measure of the gradient in field line mapping on a surface or between two surfaces \citep{Titov2002}. Quasi-separatrix layers (QSLs), defined by high $Q$ values, are known to be the preferred locations for the occurrence of reconnection, and often exhibit a complex three-dimensional structure \citep{Demoulin1996}. The intersection of two QSLs, a generalization of a two-dimensional magnetic X structure, is referred to as a hyperbolic flux tube (HFT). We investigate magnetic connectivities by tracing field lines point-wise with a fourth-order Runge--Kutta method, and use footpoint positions of field lines in the photosphere to calculate $Q$. The results of the potential field modeling and their comparison with SDO/AIA observations are shown in Figure \ref{Figure5}. We find that the overlying field of the filament system is similar to the \textit{dome-plate geometry} presented in \cite{Chen2020}. This is different from the widely accepted model for circular-ribbon flares \citep{masson2009, wang2012}, since no null point is identified around the expected location, but the large-scale structure of the field possesses considerable analogy. Underlying this analogy is the fact that the C-shaped filament channel encloses an area of (negative) polarity and is nearly completely surrounded by the opposite (positive) polarity (Figure~\ref{Figure4}). Figures \ref{Figure5}(d)--(f) show the $Q$ factor isosurfaces of the potential field, computed using the method by \citet{liu2016}. The magnetic configuration appears to be in agreement with the conjecture  by \citet{wang2014} that a quasi separator separates the dome in two parts. 

\subsection{Non-potential Field Model}

We have constructed 30 models in total, and then checked the relaxed models against the observations, in order to constrain the axial ($\Phi_\mathrm{axi}$) and poloidal fluxes ($F_\mathrm{pol}$) of the inserted flux bundles. The relaxation is performed with a similar setting to Relaxation 1 as shown in Table 2 of \citet{su2011}. The range of the parameter variations in this grid of models is the following: $\Phi_\mathrm{axi}=[2,6]\times10^{20}$ Mx; $F_\mathrm{pol}=[-1,1]\times10^{10}$ Mx cm$^{-1}$. Based on visual inspection, the best-fit model is determined by comparing the selected model field lines with the observed filament before, during, and after the eruption as shown in the first column of Figure~\ref{Figure6}. Some parameters, e.g., footpoint location and length of the inserted flux bundle, are well constrained by observations. For other parameters, there is a moderate freedom, e.g., for the most important parameter $\Phi_\mathrm{axi}$, we find a range $\Phi_\mathrm{axi}=[3,4]\times10^{20}$ Mx to be consistent with the observations. The values $\Phi_\mathrm{axi}=3\times10^{20}$ Mx (lower flux bundle) and $4\times10^{20}$ Mx (upper flux bundle) yield the overall best agreement with the data. Yet other parameters are not well constrained by the data; in particular, the relaxed configurations are similar in the whole considered range of $F_\mathrm{pol}$. 

\subsubsection{Confined Partial Eruption}

\begin{figure*}
    \centering
    \resizebox{\hsize}{!}{\includegraphics{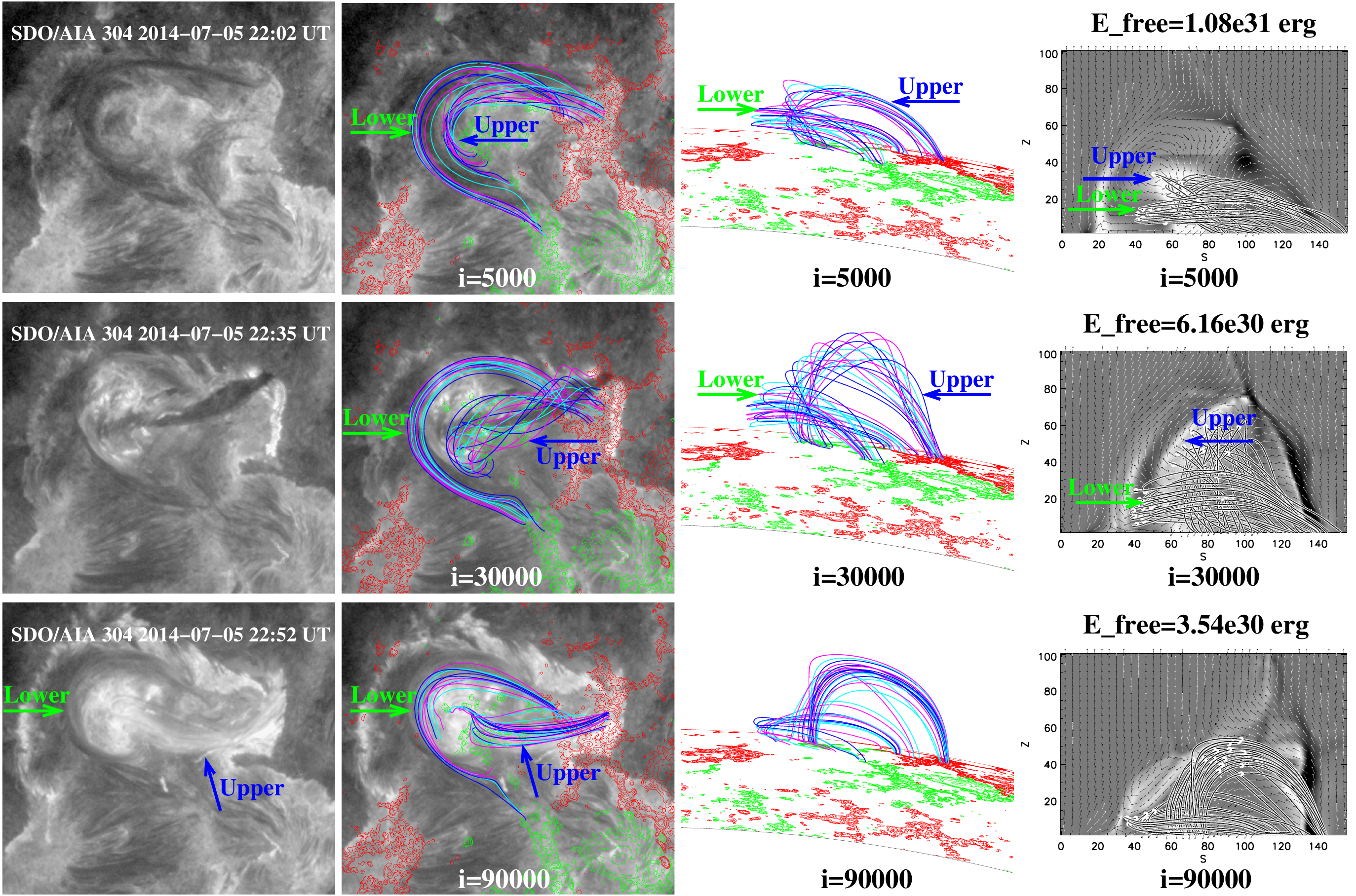}}
    \caption{The first column shows three AIA images in 304 {\AA} before, during, and after the failed filament eruption. Top and side views of selected field lines from the best-fit model after 5000, 30,000, and 90,000 relaxation iterations are presented in the second and third columns. The field lines were selected for each iteration individually. The fourth column displays the corresponding distribution of the current density (gray-scale image) and magnetic vectors (black and white) in a vertical slice at the position of the yellow line in Figure \ref{Figure4}. The intersections of the field lines with the plane of the slice are marked with white dots. The red and green contours represent the positive and negative polarities observed by HMI at 22:00 UT, respectively. The axis unit in the right column is the cell size of our model.}
    \label{Figure6}
\end{figure*}

Figures~\ref{Figure6}--\ref{Figure8} present results of the best-fit model with two flux bundles inserted along the blue paths shown in Figure \ref{Figure4}. The flux bundle FR1 connecting the footpoints L1 and L2 corresponds to the lower filament, above which is the shorter flux bundle FR2 (connecting the footpoints U1 and U2) corresponding to the upper filament, whose footpoint in the negative polarity is chosen to lie near the area of draining filament material. The initial poloidal fluxes of the inserted flux bundles are set to be $0$ Mx cm$^{-1}$, i.e., both flux bundles initially contain no twist, to minimize the influence of the insertion procedure on the resulting twist. The axial fluxes of FR1 and FR2 are $3\times 10^{20}$ Mx and $4\times 10^{20}$ Mx, respectively. 

The first column of Figure \ref{Figure6} presents three AIA images in 304 {\AA} before, during, and after the failed filament eruption. From the second to the fourth column, the individually selected field lines and the distribution of current density in the model are displayed after different numbers of magnetofrictional iterations. A comparison of the images in the first and second columns shows that the selected field lines after 5000, 30,000, and 90,000 iterations do qualitatively match the observed filaments before, during, and after the eruption. The side view of the same field lines in the third column and the current density distribution and field vectors in the cross section in the forth column show that the upper flux bundle FR2 rises to near the top of the separatrix dome from 5000 to 30,000 iterations, showing an eruption as observed. At the same time, the upper flux bundle acquires twist. The resulting state conforms to the projected position and the twisted appearance of the erupted upper filament (Figure~\ref{Figure3}f). Subsequently, the twisted flux bundle stops its rise under the separatrix dome and is largely eroded (see fourth column), which will be analyzed in detail in Section~\ref{sss:twist}. The remaining flux under the top of the dome shows an untwisted flux bundle with a slight inverse C shape (second column), also largely similar to the observed draining filament. The lower flux bundle remains nearly unchanged throughout the relaxation, which also agrees with the observations. After 90,000 iterations, it is only slightly shorter and lower than at 5000 iterations. After 5000, 30,000 and 90,000 iterations, the free magnetic energy in the best-fit model is $1.08\times10^{31}{\ }$erg, $6.16\times10^{30}{\ }$erg, and  $3.54\times10^{30}{\ }$erg, respectively, as listed in the fourth column of Figure~\ref{Figure6}. Therefore, the evolution of the observed filament eruption corresponds to the decrease of free magnetic energy. This results from the decrease of the current flowing through the upper flux rope as the rope rises and likely also from magnetic reconnection between the inserted flux bundles and the surrounding fields, which is enabled by numerical diffusion and considered in more detail below.

\subsubsection{Magnetic Topology versus Observations}

\begin{figure*}
    \centering
    \resizebox{0.9\hsize}{!}{\includegraphics{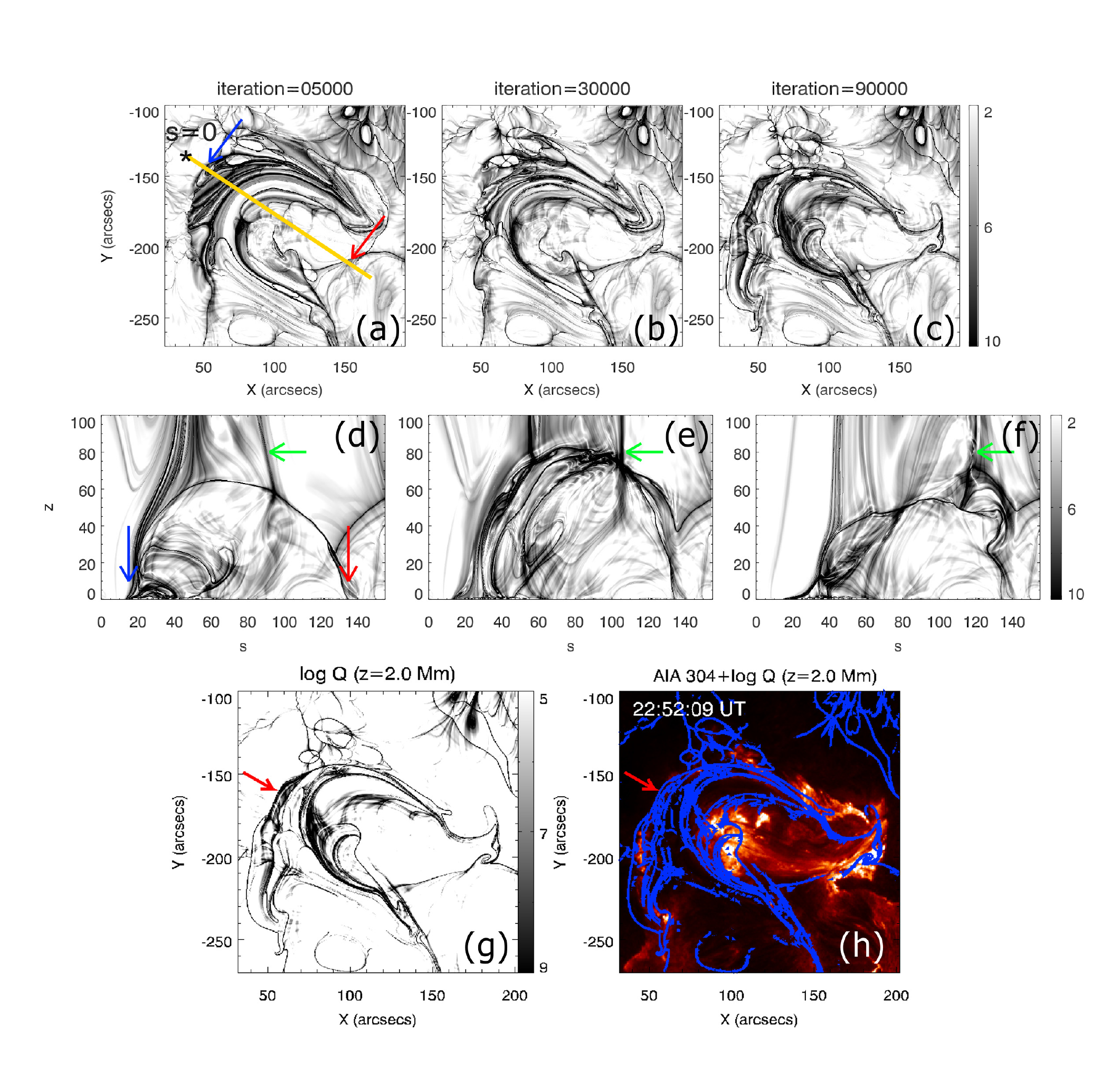}}
    \caption{Maps of $\log Q$ for the best-fit model and comparison with observations. $\log Q$ maps at a height of $z = 1.8$ Mm (top row) and vertical slices of $\log Q$ map (middle row) after different magnetofrictional iteration steps along the yellow line with the start point marked as a black asterisk, which is the same as the yellow line in Figure \ref{Figure4}. The axis unit in the middle row is the cell size of our model. Bottom row: $\log Q$ map for the model after 90,000 iterations of relaxation at a height of $2$ Mm above the solar surface (g) and the same $\log Q$ map (blue) superimposed in the image observed in 304{ \AA} by AIA (h). Green arrows in the middle row mark the plate-shaped QSL. The red arrow in the bottom row marks the high-$Q$ area on the eastern side of the inserted flux ropes.}
    \label{Figure7}
\end{figure*}

The top two rows of Figure \ref{Figure7} show the evolution of $\log Q$ in the photosphere and in the vertical cross section along the yellow line marked in panel (a). The  platelike shape extending from the top of the dome-like structure (Figure \ref{Figure5}) is the analog of the outer spine in a classical null-point topology. It is seen as the nearly straight high-$Q$ line marked by green arrows in Figures~\ref{Figure7}(d)--(f). The $\log Q$ maps show significant changes during the relaxation process. The dome-like structure, outlined as a high-$Q$ arch in the cross section, first expands and rises, and then shrinks. Simultaneously, the plate gradually moves toward the open side of the C-shaped structure, in a similar direction as the erupting filament.  

After 90,000 iterations, the $\log Q$ map at a height of $2$ Mm above the solar surface and the comparison with an image observed in $304$ {\AA} are presented in the bottom row of Figure \ref{Figure7}. Their comparison shows that the eastern outer ribbon of the main flare, R4, compares favorably with the eastern arc of high-$Q$ lines in the $\log Q$ map (marked by a red arrow). 

\subsubsection{Twist Evolution}\label{sss:twist}

\begin{figure*}
    \centering
    \resizebox{0.9\hsize}{!}{\includegraphics{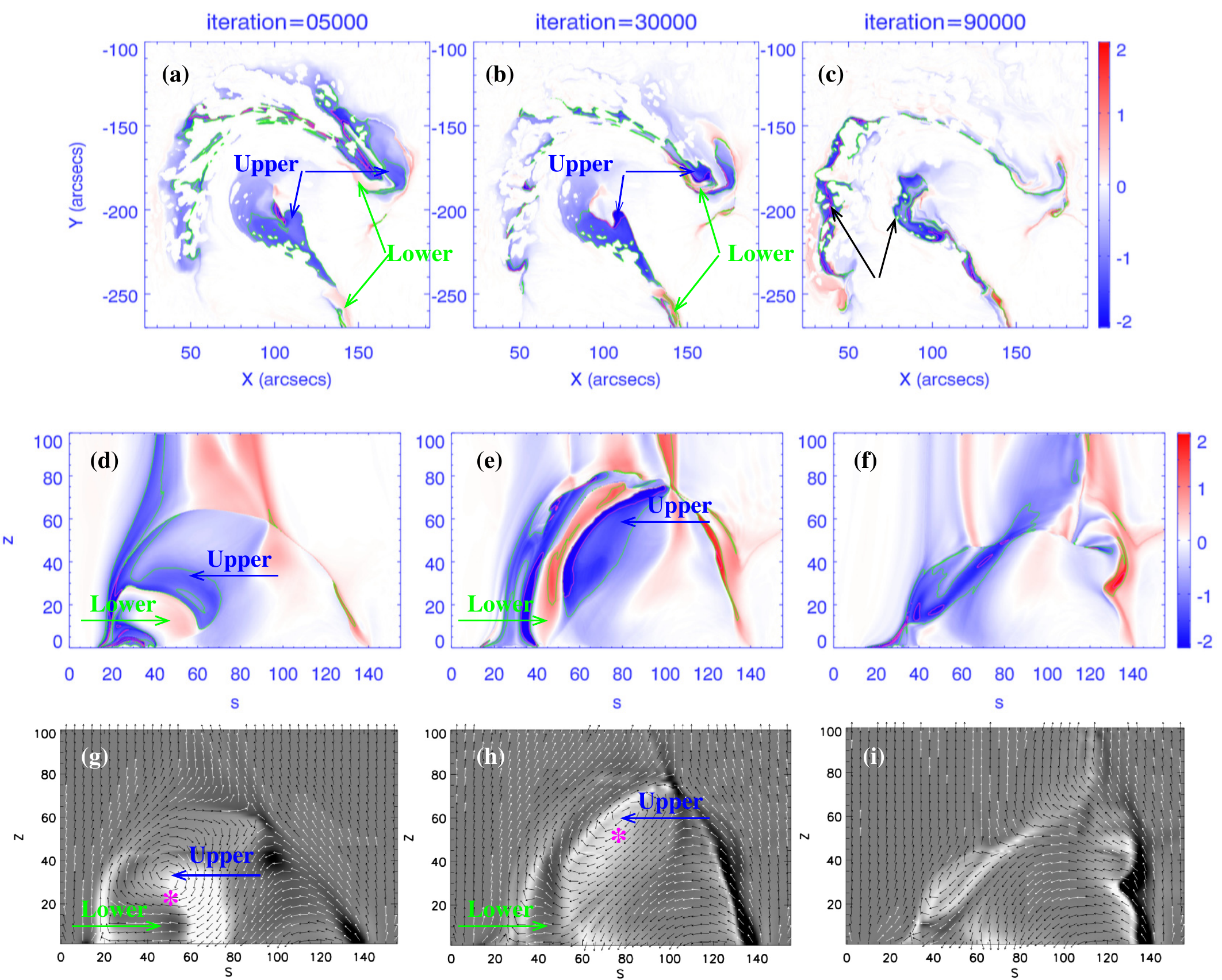}}
    \caption{Evolution of local magnetic twist $T_\mathrm{w}$ in the non-potential model. (a)--(c) Photospheric $T_\mathrm{w}$ maps after different iterations of the magnetofrictional relaxation. (d)--(f) Corresponding vertical $T_\mathrm{w}$ maps along the yellow line marked in Figure \ref{Figure4}. Green contour: $\left|T_\mathrm{w} \right| = 1.0$; Pink contour: $\left|T_\mathrm{w} \right| = 1.75$. (g)--(i) Vertical maps of the current density in the same vertical slice overlaid with in-plane field vectors. The approximate axis positions of the upper and lower flux bundle are marked with blue and green arrows, respectively. The critical height for the onset of the torus instability is shown by a magenta asterisk, in (g)--(h). The axis unit in the bottom two rows is the cell size of our model.} 
    \label{Figure8}
\end{figure*}

A flux rope is a group of helical field lines collectively winding around a common axis \citep{Priest2014, Liu2020}, which is called the magnetic axis. The winding is measured by the twist angle of the field lines about the axis, which can be expressed through the local axial and azimuthal field components $B_z(r)$ and $B_\theta(r)$, the length of the axis, $l$, and the distance from the axis, $r$, as $\phi(r)=lB_\theta/(rB_z)$, if the flux rope is cylindrically symmetric. Often, a radial average, $\phi$, is referred to as the twist of the flux rope. The twist number is $N=\phi/(2\pi)$. In practice, flux ropes are often complex structures, so that it is difficult to precisely locate the axis and perform the radial average. It is far more straightforward to compute a local twist number of each individual field line $T_\mathrm{w}=\int (\nabla \times \textbf{\textit{B}}) \cdot \textbf{\textit{B}}/(4\pi B^{2})dl$, where $dl$ is the line element along a field line \citep{berger2006}. This quantity measures how the field lines in an infinitesimally small vicinity wind about the considered field line, i.e., it does not refer to the axis of the flux rope. In the vicinity of the axis, $T_\mathrm{w}$ is close to $N$, and agrees with $N$ if the rope is axially symmetric. As an indication, moderate differences of $N$ being smaller than $T_\mathrm{w}$ by $\sim$0.5--1 have been found in two specific solar cases \citep{liu2016, Kliem2021}. If there exists a flux rope, a map of $T_\mathrm{w}$ should generally reveal its existence and location, and provide a rough estimate of its twist $\phi$. However, one has to be careful in the interpretation of such maps, because $T_\mathrm{w}$ is nonzero also in current layers, which complicates the interpretation. We map $T_\mathrm{w}$ by integrating its local density, $\alpha/4\pi$, where $\alpha=\mu_0 J_\parallel/B$ is the force-free parameter, along each field line, using the code package provided by \citet{liu2016}. The calculated $T_\mathrm{w}$ maps after 5000, 30,000, and 90,000 iterations are presented in the left, middle, and right columns of Figure \ref{Figure8}, respectively. The green and pink contours indicate the areas with local twist number $|T_\mathrm{w}|>1$ and $|T_\mathrm{w}|>1.75$, respectively. 

As mentioned above, no twist (i.e., no poloidal flux) is carried by either of the inserted flux bundles initially. After 5000 iterations, i.e., at the moment before the eruption, twist has accumulated significantly in the upper flux bundle, as shown in Figures \ref{Figure8}(a) and (d). The local  twist in the upper flux bundle $T_\mathrm{w}<-1$ (blue area; blue arrows) has its sign in agreement with the prevailing helicity in the filament channel as revealed by the erupted filament and the brightenings preceding the eruption. The area of highest $|T_\mathrm{w}|$ is close to a swirl in the in-plane field vectors in Figure~\ref{Figure8}(g), so the formation of a flux rope is indicated, with $T_\mathrm{w}$ here largely representing true twist about the axis in the center of the swirl. 

The maximum local twist of the upper flux rope reaches $T_\mathrm{w}\approx-1.75$ (roughly corresponding to $\phi \approx-3.5\pi$). This is close to the threshold of the helical kink instability \citep{fan2003, torok2004}. However, the average local twist ($T_\mathrm{w} \approx -1.45$) remains clearly lower and may be even lower if $N<T_\mathrm{w}$. Therefore, the modeling suggests that the helical kink instability did not commence, in agreement with the indication from the observations that lack clear signatures of the development of writhe. 

The local twist accumulated by the lower flux bundle is far weaker and positive ($0<T_\mathrm{w}<1$; red area and green arrows). The in-plane field vectors in Figure~\ref{Figure8}(g) do not show any indication of swirl in this area, so the local twist $T_\mathrm{w}$ here does not indicate any true twist about a magnetic axis. Overall, we find that the filament channel contains a flux rope atop a flux bundle with negligible twist. This bears some similarity to another active region analyzed in \citet{awasthi2019}, whose structure was confirmed by mass motions.

After 30,000 iterations, the area with $T_\mathrm{w} < -1.75$ has built up clearly in the upper flux rope (Figures~\ref{Figure8}(b) and (e)) and continues to enclose the magnetic axis (Figure~\ref{Figure8}(h)). The increase of the twist is also clearly seen in the field line plots in Figure~\ref{Figure6}. In contrast, the local twist number stays at low positive values, $T_\mathrm{w}<1$, in the original area of the lower inserted flux bundle. The field lines started in this area continue to show a flux bundle of negligible twist that runs along the whole C-shaped filament channel (Figure~\ref{Figure6}, middle row).  Additionally, an elongated region of positive $T_\mathrm{w}$ appears between the upper flux bundle and the dome, and acquires a higher local twist of $T_\mathrm{w}\sim1$, which is often considered to indicate a flux rope. However, there continues to be no support for the presence of a flux rope structure from the map of the in-plane field vectors (Figure~\ref{Figure8}(h)). The elongated shape and location of this relatively high-$T_\mathrm{w}$ area are instead suggestive of a current layer steepened between the dome and the rising upper flux bundle. Overall, the stable lower filament is modeled by the flux bundle of negligible twist that stays at its original height, and the erupting upper filament is modeled by a rising flux rope of increasing twist. 

A comparison of the middle and right columns of Figure~\ref{Figure8} shows a significant decrease of the twist in the upper flux bundle during the further evolution from 30,000 to 90,000 iterations. At the same time, the twist number increases significantly in the flare ribbons R4 and R5 of the main eruption (black arrows). This evolution suggests that the twist of the upper flux rope is transferred to the post-flare arcade field whose footpoints are highlighted by these flare ribbons. In these locations, the increase in the local twist $T_\mathrm{w}$ indicates the steepening of a current layer (a QSL according to Figure~\ref{Figure7}(c)), rather than the buildup of another flux rope.

The buildup of twist in the upper flux bundle is a natural result of the magnetofrictional evolution after the insertion of the flux bundle along the PIL. The ambient potential field points essentially perpendicularly to the PIL and the flux bundle. The magnetofrictional relaxation is a strongly diffusive process that creates a transition layer of smoothly changing field directions between these flux systems, with the current (which represents the free energy) mainly flowing in this layer. The field directions in the transition layer are intermediate between the axial and perpendicular directions, therefore yielding helical field lines that wrap around the inserted flux bundle, which becomes a flux rope. 

Initially, the layer forms as an annulus around each of the inserted flux bundles and can be seen in cross-sectional maps of the current density, $Q$, and $T_\mathrm{w}$. However, the layer largely diffuses away around the lower bundle in the first few thousand iterations and becomes less well defined in maps of the current density compared to $T_\mathrm{w}$ maps. This results from the present model's tendency to develop its current density in the volume of the inserted flux bundles, not only in the surface layer. The $T_\mathrm{w}$ map becomes the most sensitive diagnostic of the annulus, which is still well defined for the flux rope at 5000 iterations and still indicated at 30,000 iterations (Figures~\ref{Figure8}(d)--(e)). This magnetofrictional buildup of the current density and twist in the flux rope proceeds at scales of several $10^4$ iteration steps, i.e., continues into the rise phase of the rope (Figure~\ref{Figure6}).

In the magnetofrictional evolutionary model, the twist of the flux rope is built up in place by diffusion of the magnetic information from the inserted and the ambient flux, into the layer between them. There is no bodily transport of twisted flux into the layer. This differs from the relevant path of MHD evolution of a nearly ideal magnetofluid, which represents the conditions in the corona. In this case, twist can be transported along a flux rope or added at the surface as new layers of twisted flux. The former would occur if the flux rope is in the process of emergence \citep{Parker1974} or if a vortex would act at its footpoint(s); however, neither of these is observed in the course of the present event. The latter results from the standard flare reconnection underneath an erupting flux rope and, in the present event, is not only expected but also indicated by the bright threads wrapping around the upper filament during its rise. Despite the difference in the way the twist is built up, the underlying reason is the same, i.e., the conversion of mutual-helicity into self-helicity during the interaction of the flux rope with the ambient field. 

Table~1 lists various quantities of the best-fit model after different relaxation iterations, all computed in the HIRES region. The average degree of local twisting, $\left<T_\mathrm{w}\right>_\mathrm{box}$, is computed by integrating the absolute values of $T_\mathrm{w}$ over the bottom area of the HIRES volume and dividing this by twice the bottom area (because the local twist of each field line is summed up twice, once at each footpoint). $T_\mathrm{w}$ at the bottom plane is the value integrated over each individual field line rooted in the bottom plane, i.e., Equation~(7) in \cite{liu2016}. Therefore, the change of $T_\mathrm{w}$ in the bottom plane represents the change of magnetic twist in the relevant three-dimensional volume. \cite{Berger1984} established the relative magnetic helicity, $H_\mathrm{r}$, as a relevant physical quantity in magnetically open volumes, like the solar corona. $H_\mathrm{r}$ measures the linkage of magnetic flux tubes in the system, and is calculated following Appendix~B in \cite{Bobra2008}. Table~1 shows that all of the global non-potential measures, including the free magnetic energy divided by the potential field energy, the flux-normalized relative magnetic helicity, the average degree of local twisting of the HIRES region, and the average twist of the flux rope, decrease weakly with increasing iteration numbers. From $i$=30,000 to $i$=90,000, 27\% of $\left<T_\mathrm{w}\right>_\mathrm{box}$ and  14\% of the flux-normalized relative magnetic helicity, $H_\mathrm{r}/F_\mathrm{u}^2$, where $F_\mathrm{u}$ is the unsigned magnetic flux through the bottom plane, have decayed by numerical diffusion.

\begin{figure*}
      \begin{interactive}{animation}{Figure9-online-animation.mp4}
      \centering
      \resizebox{0.9\hsize}{!}{\includegraphics{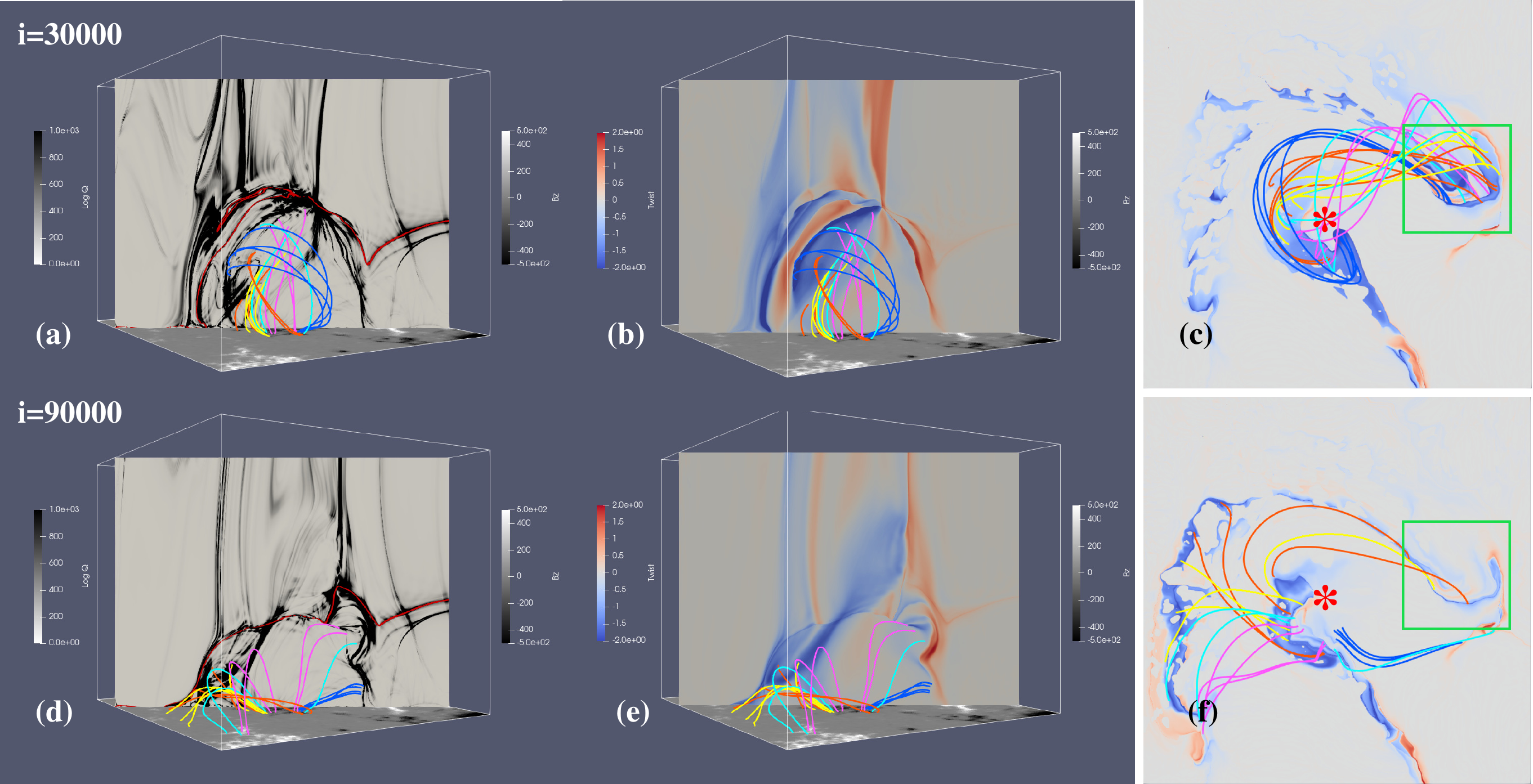}}
      \end{interactive}   
      \caption{Change of the selected field lines with fixed southern footpoints (marked with a red asterisk) from 30,000 (top row) to 90,000 iterations (bottom row). The left and middle columns show the field lines going through the vertical $\log Q$  and $T_\mathrm{w}$ maps of Figures~\ref{Figure7} and \ref{Figure8}. The bottom image in each panel shows the LOS photospheric magnetogram taken at 22:00~UT by HMI. The right panel shows the selected field lines overlaid on the photospheric $T_\mathrm{w}$ maps from the observation view angle.  An animation of the field line plots with the same field of view as the third column of this figure, covering the time interval from $i$=5000 to $i$=90,000, is available. The real time duration of the animation is 2 s.\\(An animation of this figure is available.)}
	\label{Figure9}
\end{figure*}

In order to understand the role of magnetic reconnection in the twist release of the upper flux rope, we investigate the change of the field lines in the model between 30,000 and 90,000 relaxation iterations in Figure~\ref{Figure9}, using a procedure similar to that presented by \citet{Aulanier2019}. At 30,000 iterations, we trace the field lines inside the erupted flux rope from a vertical slice to their photospheric footpoints (panels (a)--(c)). The field lines are then traced from the footpoints in the negative polarity (red asterisk), which lie near the observed ribbon R5 shown in Figure~\ref{Figure3}, in the model after 90,000 iterations (panels (d)--(f) of Figure~\ref{Figure9}). The comparison and the online animation associated with Figure~\ref{Figure9} show that these field lines of the flux rope have changed into field lines representing flare loops that connect the main flare ribbons (pink arrows in Figure \ref{Figure3}). This evolution demonstrates magnetic reconnection between the rising flux rope and ambient arcade-like flux that is rooted in the outer ribbon R4. The reconnecting flux rope field lines obtain a new footpoint in the large ribbon R4 and produce flare loops, as termed \textit{ar-rf reconnection} by \citet{Aulanier2019}. The upper flux rope begins to lose twist when this reconnection occurs near the intersection of the dome- and plate-shaped QSLs. This transfer of twist from the flux rope to the flare loops (\textit{twist release}) is clearly seen in the twist maps in Figures~\ref{Figure9}(c) and (f) as a decrease of $T_\mathrm{w}$ at the footpoint areas of the flux rope and a simultaneous increase of $T_\mathrm{w}$ at the footpoint areas of the reconnected field lines. In addition to this transfer of twist, there is a reduction of the average twist in the box, due to the numerical diffusion of the magnetofrictional scheme (Table~1).

To compare the transfer of twist by reconnection with the loss of twist by numerical diffusion, we compute the average degree of local twisting in the remote footpoint area of the flux rope, $\left<T_\mathrm{w}\right>_\mathrm{FR}$, at $i$=30,000, marked by a green box in Figure~\ref{Figure9}(c), and its change by $i$=90,000. Different from the flux in the footpoint area marked with the asterisk in Figure~\ref{Figure9}, the flux rooted in the remote footpoint area does not connect to the places that the twist is transferred to, therefore providing a measure of the twist released from the flux rope. The integral of $T_\mathrm{w}(x,y)$ in the remote footpoint area is divided by the footpoint area for the correct average. It is found that 89\% of $\left<T_\mathrm{w}\right>_\mathrm{FR}$ at $i$=30,000 is released by $i$=90,000. This is much larger than the average decrease (27\%) of $\left<T_\mathrm{w}\right>_\mathrm{box}$ by numerical diffusion in the same period. This lends support to our conjecture that a magnetofrictional description can capture some essential structural changes of a magnetohydrodynamic evolution, such as those changes due to magnetic reconnection. However, comparative studies of magnetofrictional vs. MHD modeling are needed to obtain a deeper understanding of their respective capabilities.

\subsubsection{Effect of Evolving Magnetogram}

In our simulation, the normal component of the magnetic field in the bottom plane is fixed during the magnetofrictional evolution, but the magnetic field in the solar surface does change somewhat during the relatively long-lasting eruption (i.e., from 22:02 UT to 22:52 UT). In order to investigate the effect of the magnetic field change on the modeling results, we have also constructed both a potential and a non-potential magnetic field model based on the LOS photospheric magnetogram taken at 22:52 UT and performed the corresponding topological analysis. We find that the values and shape of the high-$Q$ and high-$T_\mathrm{w}$ volumes do change, but the key features remain solid. In particular, the structure of a flux rope atop a flux bundle of negligible twist  and the buildup and release of twist of the erupting upper flux rope are essentially the same.

\section{Discussion} \label{discussion}

\subsection{Initiation of Eruption}

\citet{kliem2014} have discussed three possible trigger mechanisms for the eruption of the upper part in a double-decker filament:  (1) a weakening of the overlying field, or a transfer of flux from the lower to the upper filament; (2) a tether-cutting reconnection at the HFT between two filament flux ropes, which also transfers flux to the upper rope; and (3) the buildup of supercritical twist for onset of the helical kink instability only in the upper flux rope.  

In our case, most of the intermittent bright ejections flow along the lower filament. Only the strongest one, which leads to the precursor flare, also shows an upward displacement, in addition to its motion along the lower filament, before wrapping around the upper filament. This ejection may transfer some flux from the lower to the upper filament. Since the main eruption evolves seamlessly out of this ejection, it is possible that the ejection plays the role of the \textit{final drop}, destabilizing a structure on the verge of instability.

The same triggering mechanism can be provided by the reconnection of ambient flux between the filaments during the C1.3 precursor flare, which is suggested by the occurrence of brightened threads wrapping around the upper filament and by the ribbon pair R1 and R2 marked in Figure~\ref{Figure1}(n) (Section~\ref{ss:main_eruption}). Such reconnection is in line with the findings reported by \citet{liu2012b} and \citet{zhu2015}. 

The maximum twist angle of the upper flux rope after 5000 iterations in our model reaches $\sim{3.5\pi}$ (provided $N$ is not significantly smaller than $T_\mathrm{w}$), which is close to the threshold of the helical kink instability. However, the average twist is lower, and the observations show no signature of twist-to-writhe conversion, which is the typical telltale sign that the kink instability is involved in the destabilization of a flux rope. Therefore, the occurrence of this instability in the event is unlikely. 

One must keep in mind that the first compelling sign of evolution toward eruption, i.e., the onset of a slow westward displacement of the upper filament, suggestive of a slow-rise phase, occurs before the above potential triggers. The only associated process in the corona is the decaying filament oscillation, which fails to trigger the eruption in the present event. Therefore, the displacement may result from long-lasting,  tether-cutting reconnection driven by photospheric flux cancellation (Section~\ref{sss:cancellation}). 

In instability models for eruptions, a trigger may only push the magnetic structure across the boundary of stability, but the subsequent onset of instability drives the eruption. Therefore, we have to ask whether an ideal MHD instability could have followed the potential triggers identified above. While the helical kink is not supported, our modeling does support the occurrence of the torus instability. 

\begin{figure*}
	\centering
	\resizebox{0.9\hsize}{!}{\includegraphics{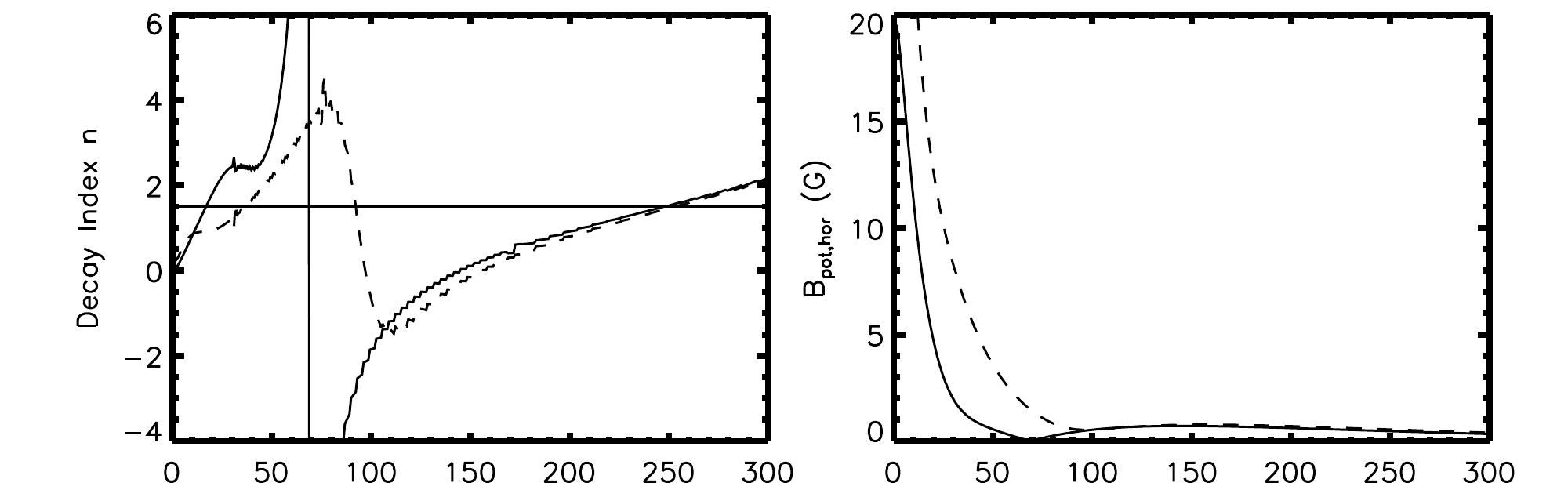}}
	\caption{Decay index height profiles (left panel) computed from the horizontal component of the potential field (right panel). The positions of the solid and dashed curves relevant at $i$=5000 and $i$=30,000, respectively, are marked as magenta asterisks in Figures~\ref{Figure8}(g) and (h), as well as in Figure ~\ref{Figure4}. }
	\label{Figure10}
\end{figure*}

The canonical value of the critical decay index for the onset of the torus instability is $n_\mathrm{cr} = 1.5$ \citep{kliem2006}, which is considered as a reference in our analysis. We find that the corresponding critical height, computed from the horizontal component of the potential field, varies significantly along the eruption trajectory. At the eruption onset, the estimated axis height of the upper flux rope (the center of the swirl in the magnetic vector distribution) of $21$ Mm exceeds the critical height for the torus instability ($16$ Mm; magenta asterisk); see Figures~\ref{Figure8}(g) and~\ref{Figure10}. This suggests that the instability initiates the upper filament eruption. Similarly, the magnetic flux rope axis in the best-fit model after 30,000 iterations has a height of $\sim{41}$ Mm, slightly above the critical height of $36$ Mm; see Figures~\ref{Figure8}(h) and~\ref{Figure10}. This is fully consistent with the status in the course of the eruption. 

As a further potential eruption mechanism, the double arc instability (DAI) has been proposed as a precursor process or as an alternative to the torus instability \citep{Ishiguro2017, Kusano2020}. This instability can result when tether-cutting reconnection merges two arc-shaped flux bundles into a double-arc shape, which implies an additional upward Lorentz force in the bent flux at the reconnection point. The data of the event suggest the occurrence of topologically similar tether-cutting reconnection at the HFT between the two filaments from the brightening of multiple sheared loops during the precursor flare (see Section~\ref{ss:precursor}, Section~\ref{ss:main_eruption}, and the animation of the AIA $171$ {\AA} data in Figure~\ref{Figure2}). However, the multiple loops represent a layer of flux wrapping around the upper filament, sharing its magnetic axis; they do not represent a new, double-arc-shaped flux bundle and current channel, which is the prerequisite equilibrium structure of the DAI. The only clear flux rope structure suggested by the observations and the modeling is the upper filament, which maintains the structure of a single arc throughout the event. On the other hand, the Lorentz force in the upper reconnected flux (visualized by the brightened multiple sheared loops) points upward and supports the subsequent eruption, similar to the action of a DAI.

\subsection{Confinement}

Various mechanisms can stop an eruption. \textbf{Observations and simulations reveal that confined eruptions may occur because the strong overlying magnetic loops form a confining cage that can prevent the escape of flux ropes} \citep{torok2005, DeVore2008, Guo2010, Amari2018, Yang2019, zheng2019}. Such a strong \textit{strapping field} above the eruption site provides a torus-stable regime above a torus-unstable one \cite[see examples in][]{WangD2017}. In addition, a flux rope is unable to erupt successfully if the tension force of a guide field is sufficiently strong \citep{kliem2014, Myers2015}. \textbf{Alternatively, confined eruptions often occur below a dome-shaped magnetic fan surface with a coronal magnetic null point.} \cite[e.g.,][]{masson2009, LiuR2018, Yang2018}. A magnetic null facilitates the onset of magnetic reconnection because oppositely directed flux exists there in close proximity. In our case, the erupting filament is surrounded by a QSL of similar dome-plate geometry (see the opposite field directions in Figure~\ref{Figure5}(b)). When the rising upper filament meets the overlying flux at the dome region (Figures \ref{Figure7}--\ref{Figure8}), magnetic reconnection between them indeed occurs (Figure~\ref{Figure9}), which erodes the flux in the rope. This changes the forces in the rope and can halt the eruption \cite[e.g.,][]{torok2005, hassanin2016}.

Our analysis indicates that the confinement of the erupted upper filament in the investigated event can be due to a combination of the aforementioned two main mechanisms. Certainly the reconnection with overlying flux is important, as demonstrated by the draining of the filament material in the observations and as corroborated by the model. 

The observations and the potential field computation indicate that the tension force of the overlying flux is also important, providing a \textit{confining cage}. The animation of the AIA data (Figure~\ref{Figure3}) shows that the first signs of the reconnection with ambient flux---the formation of new connections in the area of the original negative footpoint---occur only after 22:39~UT. At this time, the filament has already nearly reached its final position. This indicates that the filament is strongly decelerated by the downward tension force of overlying flux, and the deceleration process is also in agreement with the large-scale structure of the overlying flux as revealed by the potential field. Figure~\ref{Figure10} shows that the flux above the QSL dome provides an extended height range where $n<3/2$. A rising flux rope can be stopped in such a torus-stable height layer if the layer contains a sufficient amount of flux \citep{DeVore2008}. We expect that this happened in the investigated event. The two flux systems pressing against each other are also likely to reconnect, and this is indeed indicated by the observations during the draining of the stopped filament. On the other hand, in the magnetofrictional evolution of the model flux rope, reconnection with overlying flux is the dominant process that stops the rise, due to the high diffusion inherent in this approximation.

The torus-stable height layer in the present active region has also been found by \citet{Filippov2020}. Assuming that a strapping field of opposite direction above a magnetic null would exert an upward Lorentz force on the rising flux rope, this author concluded that ``the most reasonable force that terminate the filament ascending above the null point is the gravity force.'' This would be valid in vacuum fields. However, in MHD, a flux rope rising into a region of oppositely directed strapping field will never penetrate into this flux; thus, it will not feel such a Lorentz force. Rather, the rope will either stretch this flux upward, creating a downward magnetic tension force \citep{DeVore2008, Kliem&al2016}, or will reconnect with this flux, which erodes the flux rope, weakening its upward Lorentz force \citep{torok2005, hassanin2016}.

\section{Summary and Conclusions} \label{summary}

We investigate the failed partial eruption of a C-shaped double-decker filament system, which occurred in AR~12104 in the southern solar hemisphere on 2014 July 5. The filaments enclose a region of negative flux that is nearly completely surrounded by positive flux. Before the eruption, a sequence of interrelated activities are observed: intermittent bright ejections, oscillation of the upper filament, and a C1.3 precursor flare. The eruption of the upper filament is associated with the main C2.5 flare with a C-shaped flare ribbon surrounding the stable lower filament and a compact ribbon forming in the enclosed polarity. The upper filament displays a prominent twisted structure when it is undergoing a failed eruption and rolls over toward the open side of the C shape during the eruption. The rise of the filament then halts; it transforms to a reversed-C shape and subsequently drains, partly to places different from its original footpoints. 

A careful examination of the photospheric magnetic field suggests that the intermittent ejections are due to flux cancellation between two small converging polarities near the northwestern footpoint of the filament. Long-lasting filament oscillations are often observed before the onset of solar eruptions; therefore, these oscillations were proposed as a precursor leading to eruption \citep{Chen2008}. However, the filament oscillation that lasts about 2 hr in the present case does not directly lead to an unstable state of the system because it completely decays about an hour before the onset of the precursor flare.

The double-decker structure of the filament system, i.e. the trapping of the filaments in two separate magnetic structures, is indicated by the different dynamic behavior of the filaments during both the precursor activities (intermittent ejections disturbing only the lower filament; Section~\ref{sss:intermittent}) and the main eruption (reconnection between the filaments, producing bright threads between them; Section~\ref{ss:main_eruption}). This is further supported by the successful model that employs a double-decker structure.

To understand how the eruption initiates and why it fails, we construct magnetic field models using the flux rope insertion method that involves magnetofrictional relaxation. The magnetofrictional evolution has no intrinsic time scale, as the velocities used in the magnetofrictional scheme are not real fluid velocities. However, the spatial structures, including the magnetic topology, are realistic and even data constrained; thus the resultant magnetic configuration can help understand the observations \citep{Cheung2012, Savcheva2015, Savcheva2016, Guo2016}. 

Topological analysis of the potential field shows that the overlying field of the filament system possesses a QSL of dome-plate geometry but without an embedded magnetic null point. A high-$Q$ plate separates the dome into two parts. Our modeling of the non-potential coronal field reproduces the observed failed eruption of the upper filament, with the lower filament remaining essentially undisturbed. This includes the accumulation and subsequent release of twist in the upper filament during the eruption. The magnetic configurations after 5000, 30,000, and 90,000 magnetofrictional iterations rather closely match the observed filament system before, during, and after the eruption, respectively. The evolving model also shows a decrease of free magnetic energy, as required for an eruption. 

After 5000 iterations, the magnetic field evolves into a configuration with a flux rope atop a flux bundle with negligible twist, which appears to closely match the observed double-decker filament at the eruption onset. Although both flux bundles were inserted without an azimuthal field component, the upper one acquires twist. The twist continues to increase until the termination of the rise after 30,000 iterations. Consistent with the observations, the twist in the upper flux rope shows a significant decrease subsequently. At the same time, the non-potentiality indicated by the photospheric $T_\mathrm{w}$ map increases substantially in the area of the main flare ribbons, suggesting a twist or current transfer from the erupted flux rope to the surrounding field, which results from the reconnection of the flux rope with the surrounding magnetic field at the high-$Q$ dome. The ribbons take the form of an extended C on the outer side of the filament channel and of a compact area inside, respectively. The C-shaped outer ribbon coincides with lines of high squashing factor $Q$ in the photosphere, similar in shape to the footpoint area of the magnetic dome in a null-point configuration. The compact inner ribbon forms at the periphery of the erupting flux rope of the upper filament, as indicated by a developing dimming. This is located near the footpoint area of the magnetic spine.

Previously, it has been found that the magnetofrictional relaxation process is capable of emulating the observed evolution of an eruption in the early phase \citep{su2011, Savcheva2016, liutie2018}. For the first time, this study shows that the magnetofriction method can also emulate qualitatively emulate the confined partial eruption and the observed  twist evolution (due to the reconnection with overlying flux) in a filament. However, cautions must be taken in applying the method and in interpreting the modeling result. For instance, because the method cannot describe any MHD waves (fast mode, Alfven mode, nor slow mode waves), it works as a reliable approximation of the MHD evolution only when the dynamics are well subsonic and sub-Alfvenic, in particular, for quasi-static evolutions. When eruptive dynamics with strong acceleration of plasma are considered, a careful comparison between the  magnetofrictional evolution and the observational data or an MHD model is still required for a representative number and range of case studies. Our magnetofrictional modeling yields magnetic reconnection of erupting flux in qualitative agreement with the observed twist release from the erupting flux and the brightening of the main flare ribbons, but it does not yield a long, thin current sheet between the erupting and overlying flux systems \citep{Guidoni2016}, similar to previous applications of the method \citep{su2011, Savcheva2016, liutie2018}. This suggests that such modeling yields the large-scale consequences but not the detailed dynamics of reconnection. 

The main results of our observational and modeling analysis of the failed eruption can be summarized as follows: 
\begin{enumerate}
\item The double-decker filament is structured as a flux rope atop a flux bundle with negligible twist. 
\item Prior to the eruption, one of the strongest intermittent ejections from an area of flux cancellation in the filament channel appears to transfer flux into the upper filament, which weakens its stability. Additionally, magnetic reconnection of the ambient field in the space between the two filaments, indicated by brightened threads immediately following the precursor flare, transfers further flux to the upper filament and simultaneously reduces the constraint imposed by the overlying field. Both mechanisms contribute to the observed slow rise and destabilization of the filament, which eventually erupts, once reaching the critical height of the torus instability. On the other hand, the $\sim{2}$ hr filament oscillation before the eruption occurs in a stable state of the system is unlikely to trigger the eruption.  
\item Both observations and magnetic modeling indicate that the erupted flux is considerably twisted, but the twist does not reach the threshold of the helical kink instability. The filament threads trace preexisting twist, which becomes apparent during the expansion and disintegration of the filament. Additional twist is added to the erupting flux in the course of the rise as a result of tether-cutting reconnection of ambient flux under the rising flux. Once reaching the overlying dome-shaped QSL, the twist of the upper filament is transferred to the post-flare arcade field whose footpoints are highlighted by the flare ribbons surrounding the lower filament. The transfer results from magnetic reconnection with the overlying field. Both the rise of the upper filament and its subsequent untwisting lower the free magnetic energy of the configuration.
\item The confinement of the eruption is caused by one or both of the following effects. (a) Motion of the erupting flux from a torus-unstable to a torus-stable height range, the latter of which is located just above the intersection of the dome- and plate-shaped QSLs. (b) Erosion of the erupted flux rope by reconnection with overlying flux at the dome-shaped QSL. This is suggested by the draining of the filament to partly new footpoint locations, which immediately follows the termination of its rise, and directly observed in the model. 

\end{enumerate}

\acknowledgments 

The authors thank the SDO and the GONG teams for providing the valuable data. We thank Jun Chen for kindly offering the code package for calculating $T_\mathrm{w}$ and the $Q$ factor and Drs. Yang Guo and Xin Cheng for valuable discussions. We also gratefully acknowledge constructive comments by the referee which have stimulated a deeper analysis, especially of the reconnection in the model and of the differences in the magnetofrictional and MHD descriptions of eruptive processes. This work is supported by the following Chinese foundations: the National Natural Science Foundation of China (NSFC; 41761134088, 11473071, 11790302, 11790300, 11773079, 41774150, 11925302, and U1731241),  and the Strategic Priority Research Program on Space Science, CAS (grant No. XDA15052200 and XDA15320301). B.K. acknowledges support by the DFG in the joint Sino-German program NSFC41761134088/DFG~KL817.8-1.

\vspace{5mm}
\clearpage

\bibliographystyle{aasjournal}  
\bibliography{ref}                

\begin{table}
\tabletypesize{\scriptsize}
\begin{center}
\caption{The Evolution of Free Energy, Helicity, and Local Twist}
\begin{tabular}{ccccc}
\tableline\tableline
             & $E_{\mathrm{free}}$/$E_{\mathrm{poten}}$ & $H_{r}/F_{u}^{2}$ & $<T_\mathrm{w}>_{\mathrm{box}}$ & $<T_\mathrm{w}>_{\mathrm{FR}}$\\
Iteration &  ($10^{-3}$) &($10^{-4}$) & ( 10$^{-2}$) & (10$^{-2}$)\\
\tableline
5000 &  9.77  &  -3.89 & 7.39  & -24.87 \\
30,000 & 5.60  &  -3.86 & 5.29 & -20.49 \\
90,000 & 3.22  &  -3.32 & 3.84 & -2.32 \\
\tableline 
\end{tabular}
\tablenotetext{}{\textbf{Note.} Global non-potential measures including: magnetic free energy ($E_{\mathrm{free}}$) divided by the potential field energy, flux-normalized relative magnetic helicity ($H_{r}/F_{u}^{2}$), average local twist of the HIRES region ($<T_\mathrm{w}>_{\mathrm{box}}$), and average twist of the flux rope of the best-fit model ($<T_\mathrm{w}>_{\mathrm{FR}}$) after different iterative relaxations. The potential field energy is $1.1\times10^{33}$ erg.}
\end{center}
\end{table}

\end{document}